\documentclass[twocolumn,amsmath,floatfix,prb,aps]{revtex4-1}
\usepackage{color}
\usepackage{amsmath}
\usepackage{pifont}   
\usepackage{graphicx} 
\usepackage{dcolumn}  
\usepackage{bm}       
\usepackage{amsfonts} 
\usepackage{amssymb}  
\usepackage{multirow} 
\usepackage{siunitx}

\usepackage[english]{babel}
\usepackage[autostyle, english = american]{csquotes}
\MakeOuterQuote{"}

\usepackage{placeins}

\usepackage{braket} 

\usepackage{bbm}

    \renewcommand{\v}[1]{\bm{\mathrm{#1}}}

\begin{document}

\title{Electronic origin of x-ray absorption peak shifts}

\author{S. Shallcross$^1$}
\author{C. v. Korff Schmising$^1$}
\author{P. Elliott$^1$}
\author{S. Eisebitt$^1$}
\author{J.~K. Dewhurst$^2$}
\author{S. Sharma$^1$}
\email{sharma@mbi-berlin.de}
\affiliation{1 Max-Born-Institute for Non-linear Optics and Short Pulse Spectroscopy, Max-Born Strasse 2A, 12489 Berlin, Germany}
\affiliation{2 Max-Planck-Institut fur Mikrostrukturphysik Weinberg 2, D-06120 Halle, Germany}

\date{\today}

\begin{abstract}
Encoded in the transient x-ray absorption (XAS) and magnetic circular (MCD) response functions resides a wealth of information of the microscopic processes of ultrafast demagnetisation. Employing state of the art first principles dynamical simulations we show that the experimentally observed energy shift of the L3 XAS peak in Ni, and the absence of a corresponding shift in the dichroic MCD response, can be explained in terms of laser induced changes in band occupation. Strikingly, we predict that for the same ultrashort pump pulse applied to Co the opposite effect will occur: a substantial shift upwards in energy of the MCD peaks will be accompanied by very small change in the position of XAS peaks, a fact we relate to the reduced $d$-band filling of Co that allows a greater energetic range above the Fermi energy into which charge can be excited. We also carefully elucidate the dependence of this effect on pump pulse parameters. These findings (i) establish a electronic origin for early time peak shifts in transient XAS and MCD spectroscopy and (ii) illustrate the rich  information that may be extracted from transient response functions of the underlying dynamical system.
\end{abstract}

\maketitle

The ultrafast control over magnetic solids by light has opened new vistas in fundamental condensed matter physics\cite{Bigot2009,Kirilyuk2010,Battiato2010,Radu2011,Eschenlohr2013,bovensiepen2009} that, with the recent introduction of first principles simulations\cite{dewhurst2018,Jin2018,Kimel2020} capable of very accurately treating the early time regime (less than a few hundred femtoseconds), can now be explored in tandem by theoreticians and experimentalists\cite{Siegrist2019,hofherr2020}. Experimentally, the measurement of dynamically evolving magnets on femtosecond to picosecond time scales rests upon spectroscopic probes, with two of the key methods transient x-ray absorption spectroscopy (XAS) and transient magnetic circular dichroism (MCD)\cite{Stamm2007a,Boeglin2010,Thole1992,Carra1993,Stamm2010,Bergeard2014,Higley2019,Hennecke2019a,Hennes2020a}. In these techniques a laser pumped magnetic material is perturbed with helicity dependent (i.e. left and right circularly polarised) light, the frequency of which resonates with the core level of one of the constituent atoms, and the response then recorded as a function of time. The normalized difference of these two helicity dependent response functions is proportional to the MCD, and represents the magnetic response of the material, while their sum is proportional to the XAS, and represents the charge response of the material. 

Integrating the MCD and XAS spectra via the sum rules\cite{chen1995,Thole1992,Carra1993} yields species resolved magnetic moments, a procedure that is rigorous for atoms and, while less rigorous, represents a "tried and tested" method for probing magnetism in solids, both in the ground state as well as in laser induced dynamical states. It is clear, however, that a wealth of information of the underlying dynamical processes are encoded in transient response functions\cite{Guyader2021,Willems2020a,Higley2019,Hennes2020a,Rosner2020}, with the extraction of magnetic moments representing only one possibility. A crucial task at the frontier of ultrafast spectroscopy is therefore interpreting the temporal evolution of spectral structures such as peak centres and widths: what insight into the underlying microscopic physics of ultrafast demagnetisation can be extracted from such spectral changes?

In laser pumped Ni it was observed that in addition to the reduction in amplitude of the MCD and XAS peaks, well known signatures of laser induced demagnetisation, the XAS peak shifts noticeably to lower energies \cite{Stamm2007a} with, in contrast, the MCD peak exhibiting no such shift \cite{Kachel2009, Higley2019, LeGuyader2022}. While the reduction in peak amplitudes can be understood in terms of the excitation of charge and a concomitant high rate of spin-orbit induced transitions from majority to minority, physics captured by the sum rules, the XAS peak shift, and the absence of such a shift in the MCD spectrum, has defied microscopic analysis. Connecting features of spin and charge response functions to underlying microscopic processes is in general a difficult task, and is made more so in the highly non-equilibrium situation of laser induced demagnetisation by the several distinct demagnetization processes that occur. This is reflected in the rather disparate set of explanations that have been put forward for the XAS peak shift in Ni: thermal effects\cite{Kachel2009}, magnon excitation\cite{LeGuyader2022}, and band mirroring\cite{Higley2019} have all been considered. Theoretically, response functions derived from the static picture with redistribution of charge were found not to be able to explain this difference between the behaviour of the XAS and MCD peaks\cite{Carva2009}.

In the present work, by performing fully quantum-mechanical state-of-the art dynamical calculations\cite{dewhurst2020,dewhurst2022,sharma2022}, we resolve this existing controversy concerning the interpretation of XAS and MCD spectral shifts in Ni. We find, in agreement with experiment, a shift downwards in photo energy of the XAS peak with almost no corresponding shift upwards of the MCD peak, moreover we predict that in Co pumped with an ultrashort laser pulse, remarkably, this effect is inverted: a large shift of the MCD peaks towards higher photon energies is accompanied by a very small shift of the XAS peak. The origin of both these effects we locate in specific features of the electronic spectrum near the chemical potential, with the increased $d$-band filling of Ni as compared to Co playing an important role. The XAS and MCD peak shifts in Ni and Co are thus shown to be a purely electronic effect resulting from laser induced changes in band occupation, providing an example of how XAS and MCD response functions may yield insight into the underlying spin dynamics beyond that provided by their use in the determination of dynamical spin moments.


\emph{Computational details}: To calculate the spin dynamics of laser pumped Ni and Co we have used the fully {\it ab-initio} state-of-the-art fully non-collinear spin-dependent version \cite{krieger2015,dewhurst2016} of time dependent density functional theory (TDDFT)\cite{RG1984}. Note that the presence of spin-orbit coupling breaks the SU(2) spin symmetry and mixes both spin channels, in turn requiring the fully non-collinear treatment of spin that we employ. In the present work we have used adiabatic local density approximation for the exchange-correlation potential and all calculations are performed using the highly accurate full potential linearized augmented-plane-wave method\cite{singh}, as implemented in the ELK\cite{elk,dewhurst2016} code. A face centred cubic unit cell with lattice parameter of $3.21\si{\angstrom}$ was used for Co, while for Ni a lattice parameter of $3.53\si{\angstrom}$ was used. The Brillouin zone was sampled with a $20\times 20\times 20$ k-point mesh for both materials. For time propagation the algorithm detailed in Ref.~\onlinecite{dewhurst2016} was used with a time-step of $2.42$ atto-seconds. The final magnetization value converges with the above mentioned computational parameters to 1.67$\mu_{\rm B}$ for Co and 0.61$\mu_{\rm B}$ for Ni.

In order to calculate transient XAS and MCD spectra for laser pumped Ni and Co at various times, the linear response formalism of the TD-DFT is the used\cite{RG1984,my-book,sharma2011}. This method\cite{dewhurst2020} is equivalent to probing the system without accounting for the width of the probe-pulse. Despite this we found that the transient response functions are in excellent agreement with experiments\cite{clemens2020,dewhurst2020}. 
In calculating the response functions we use 23 {\bf G}-vectors to capture local field effects, the bootstrap kernel to treat excitonic effects, and a smearing of 0.9~eV that is derived from the $GW$ calculated average width of the semi-core $p_{3/2}$ and $p_{1/2}$ L peaks. The position of the 2$p$-states were also determined via the $GW$ method, with the scissors corrected KS states used in calculating the response functions. The $GW$ calculations were performed at a temperature of 500~K using the spin-polarized $GW$ method\cite{gw}. The spectral function on the real axis is constructed using a Pade approximation with spin-orbit coupling was included and a Matsubara cutoff of 12~Ha used. We find that the static XAS and MCD spectra calculated with these parameters reproduces previous theoretical simulations\cite{ebert1996,wu1994} and is in good agreement with experiments\cite{chen1995,Kachel2009}.


\begin{figure}[t!]
\includegraphics[width=\columnwidth, clip]{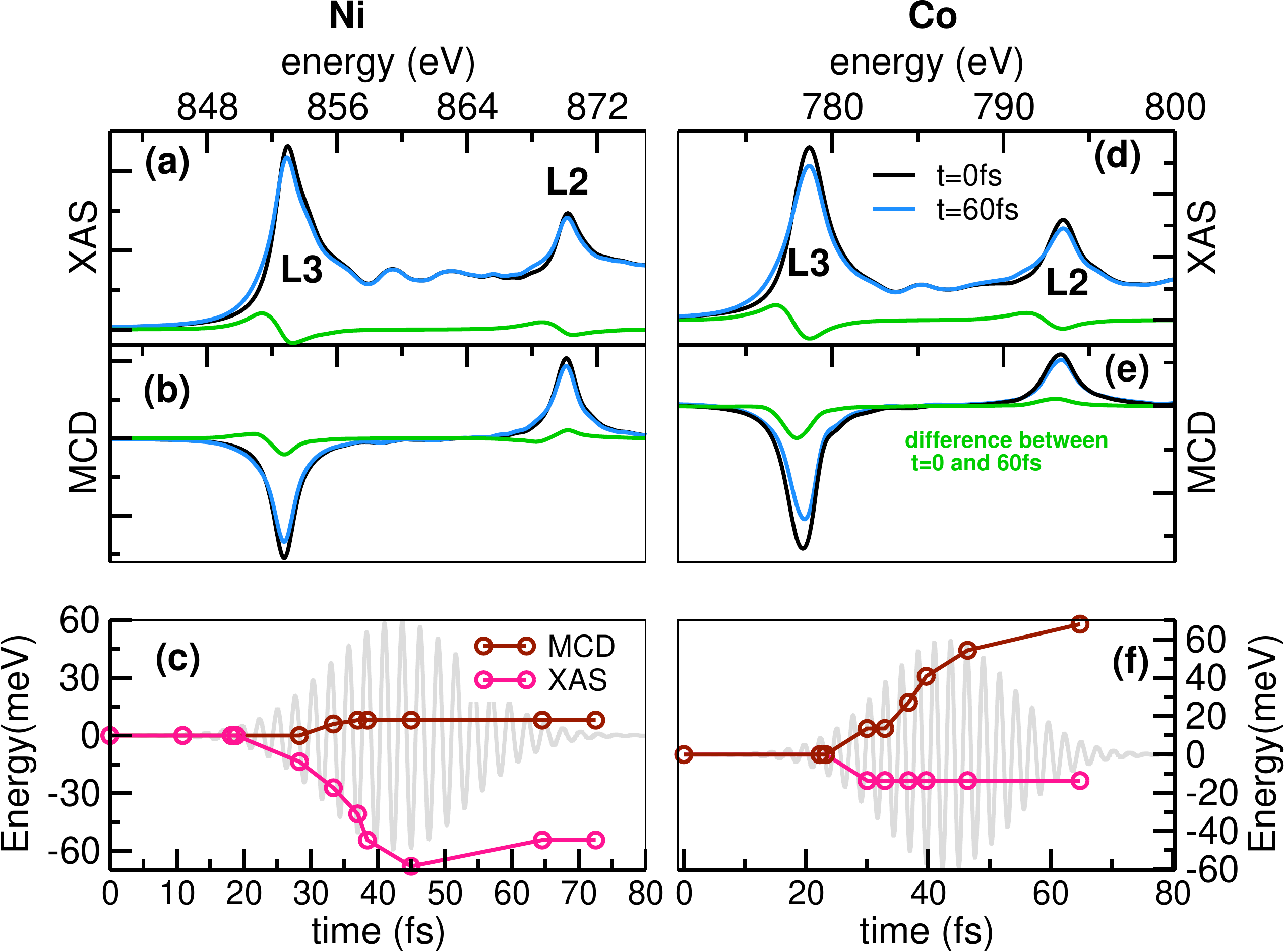}
\caption{\emph{XAS, MCD and L edge peak shifts for Ni and Co}. Shown in panels (a) and (b) respectively are the XAS and MCD spectra before and after ($t=60$~fs) laser pumping for Ni, with the difference shown in green. The energy shift of the XAS and MCD L3 peak (on meV) as a function of time is shown in panel (c), revealing a substantial shift down in energy of the XAS peak but almost no shift in the MCD L3 peak. Panels (d-f): corresponding data for Co. In dramatic contrast to Ni, for this material it is the MCD that shows a substantial shift (upwards) in energy. In both cases the same 1.55~eV linearly polarized pump pulse (with polarization parallel to the spin-quantization axis) was used with fluence 6.7~mJ/cm$^2$, full width half maximum 24.5~fs, and intensity 7.2x10$^{12}$~W/cm$^2$, the {\bf A}-field of this pump pulse is shown in grey in panels (c,f).
}\label{fig:2}
\end{figure}

\emph{Transient XAS and MCD:}  It is well established in experiment that pumping with laser light leads to demagnetization of both Ni and Co. This change in the moment can be determined by integrating the MCD and XAS spectra and using the sum rules\cite{chen1995,Thole1992,Carra1993}. However, this requires fully energy and time resolved spectra, which is experimentally very difficult to determine. The signature of this loss in spin moment also can be determined by just looking at the reduction in the peak (L3 and L2) height in XAS and MCD spectra. For Ni this reduction can be experimentally observed in the MCD spectra. Together with this reduction, the XAS (L3 edge) shows a substantial shift to lower energy, while the MCD spectra does not shift and several scenarios have been proposed to explain this\cite{Kachel2009,LeGuyader2022,Higley2019}.

To examine this in Fig.~\ref{fig:2}a,b we show the calculated XAS and MCD response function before ($t=0$) and after ($t=60$~fs) laser pumping along with their difference. A linearly polarized pump pulse (with polarization vector parallel to the spin-quantization axis) of fluence 6.7~mJ/cm$^2$, duration 24.5~fs, and a central frequency of 1.55~eV is used. From examination of the XAS it can be discerned that the L3 peak at 60~fs has shifted to lower energies. Furthermore, the weight of the XAS is also shifted to lower energies. The MCD spectra at the L3 edge merely broadens and reduces in amplitude. The difference of the XAS signal between $t=0$ and $t=60$~fs exhibits a signature bipolar behaviour-- decrease in XAS at L3 and L2 edges is accompanied by an increase at energies below L3 and L2 edges.  This is expected as the total charge is conserved, so that a decrease at the resonance must be accompanied by an increase at lower energies. 
Often this bipolar behaviour is inferred as an indication of the shifting of the peak to lower energy, however this is generally not correct, and such bipolar behaviour indicates only a shifting of the spectral weight to lower energy which might not be accompanied by a shift of the peak centre.
To unveil the dynamics of the XAS and MCD peak shifts in Fig.~\ref{fig:2}c we plot the peak centres as a function of time. Exactly as experiments report we see a substantial shift of the XAS L3 edge peak, that begins at the start of the pulse and is largely complete at the maximum of the pump pulse envelope. Again as seen in experiment the MCD peak does not show a significant shift in position with a maximum shift of +8~meV to higher energies, approximately an order of magnitude smaller than the XAS peak shift downwards in energy.

\begin{figure}[t!]
\includegraphics[width=\columnwidth, clip]{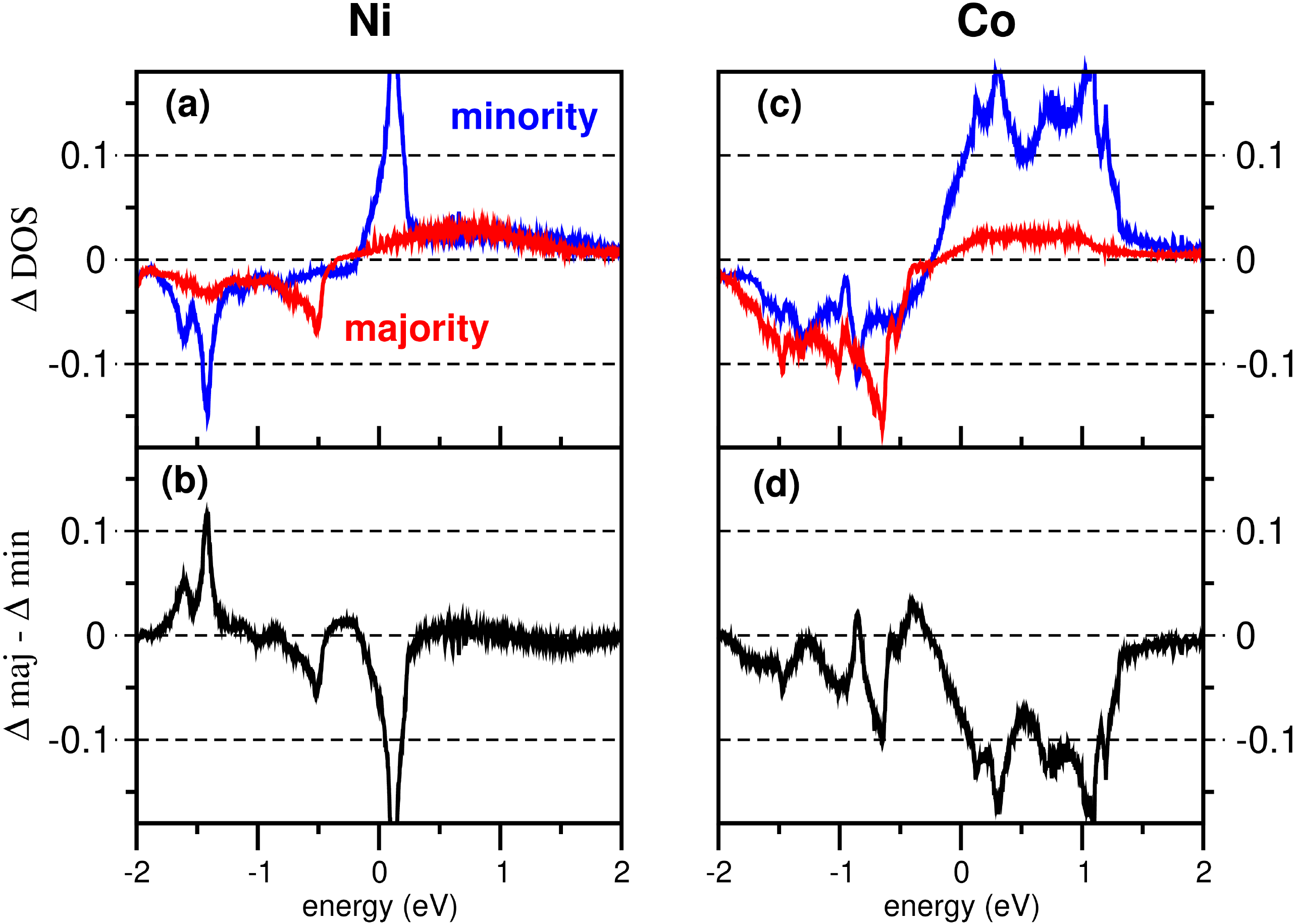}
\caption{Difference in the spin and $d$-state projected DOS (in states/eV/spin) before and 60~fs after pumping for (a) Ni and (c) Co. The energy resolved change in moment (in $\mu_B$) from the ground state to 60~fs after pumping for (b) Ni and (d) Co. The pump pulse is linearly polarized with polarization direction parallel to the spin-quantization axis. In both cases the same 1.55~eV pump pulse was used with fluence 6.7~mJ/cm$^2$, full width half maximum 24.5~fs, and intensity 7.2x10$^{12}$~W/cm$^2$.
}\label{fig:emag}
\end{figure}

Since the L-edge XAS is a direct probe of the available $d$-states and MCD of the spin-projected available $d$-states for transition from the exchange-split 2$p$-states\cite{dewhurst2022,sharma2022}, we now take a closer look at the transient occupied density of states (DOS)\cite{Elliott2016}. In Fig. \ref{fig:emag}a we present the difference in the transient occupied $d$-band DOS for Ni after the pump pulse (60~fs) and occupied $d$-band DOS of the ground state. Note therefore that in this figure positive $\Delta$DOS indicates the laser induced occupation of states, and negative $\Delta$DOS the creation of empty $d$-states. A striking feature can immediately be observed in the narrow energy range of minority channel excited charge just above $E_F$. This $\sim$0.5~eV region of laser excited charge contrasts with the situation below $E_F$ in which empty states of minority and majority charge have been created over a wide energy range of $\sim$2~eV. These empty states are created both due to laser excitation as well as due to spin-orbit mediated spin-flips. The origin of the peak shift to lower energies is now clearly revealed: as the XAS response function measures transitions from the 2$p$ states to unoccupied valence states, above $E_F$ the response function will be reduced in a narrow energy window, while below $E_F$ the response will be increased in a wide energy range. Taken together, this can only result in a shift to lower energies of the L3 (as well as L2) peak centres and spectral weight.

For elucidating the non-shifting of the MCD peak the relevant quantity is the difference of the two $\Delta$DOS (majority and minority), i.e. the energy resolved change in moment from the ground state. This change in moment is shown in Fig.~\ref{fig:emag}b for $t=60$~fs (after the pump pulse), revealing that this quantity shows two almost symmetrically (in energy) placed peaks around $E_F$ (0.2~eV above and 0.5~eV below $E_F$), however, the peak above  $E_F$ is much larger in magnitude (note the opposite sign of the peak simply follows from the fact that we plot in $\Delta$DOS the difference of majority and minority DOS, and the system has suffered an overall loss of moment). Following the same argumentation of the XAS response function, this would be expected to lead to a shift in the MCD peak to higher energies and a broadening of the peak with spectral weight shifting also to lower energies (due to a peak 0.5~eV below $E_F$). However, as the peak above $E_F$ resides in very narrow energy window the shift will be small. This is exactly what is seen in Fig. \ref{fig:2}c. In experiment this small shift is probably not resolvable and thus these results are consistent with the reported non-shifting of the MCD peak.

It is clear that the laser induced changes of occupation revealed in Fig.~\ref{fig:emag} represents a complex dynamical change in band occupation that is highly asymmetric about $E_F$, occurs over a wide range of energies, and is material specific. A simplistic model in which the occupation function is symmetrically manipulated will hence not be expected to capture the underlying physics of peak shifts\cite{Carva2009}, which requires a fully dynamical quantum-mechanical description for their explanation.

Having established that the L3 edge peak shift in Ni has an electronic origin, it is of interest to examine the corresponding spectral feature in neighbouring Co. In Fig.~\ref{fig:emag}c,d we show the change in DOS for both spin channels and the energy resolved change in moment for Co. We employ the same pulse parameters as used for Ni and again show the change in $d$-DOS and energy resolved moment between 60~fs (after the pulse) and the ground state. 

Comparison with Ni reveals a strikingly different behaviour occurs in laser pumped Co. Whereas Ni exhibited a minority peak restricted to a narrow 0.5~eV range above $E_F$, for Co minority excitations occurs over a broad range in energy: both minority and majority are excited from a 2~eV range below $E_F$ into the minority states in a similarly broad energy range of up to 2~eV above $E_F$. The L edge response is thus not expected to exhibit as pronounced a shift in its central peak as in the case of Ni. Turning to the XAS spectra for Co in Fig.~\ref{fig:2}d and the corresponding peak shift plotted in Fig.~\ref{fig:2}f we see this is exactly what occurs: the L3 peak centre shifts by a very small value of 20~meV. 

In contrast to the charge excitation the energy resolved change in moment from the ground state shows a significant skew about $E_F$ (see Fig.~\ref{fig:emag}d).
This should then result in a pronounced decrease of the leading edge of the L3 of the MCD peak, resulting in an shift of these MCD spectra to higher energies. Again, inspection of the MCD spectra of Co in Fig.~\ref{fig:2}e and the corresponding peak shift plotted in Fig.~\ref{fig:2}f reveals exactly the behaviour expected on the basis of this underlying electronic rearrangement of the moment: a dramatic shift upwards in energy of the MCD peak by 60~meV.

\begin{figure}[t!]
\includegraphics[width=\columnwidth, clip]{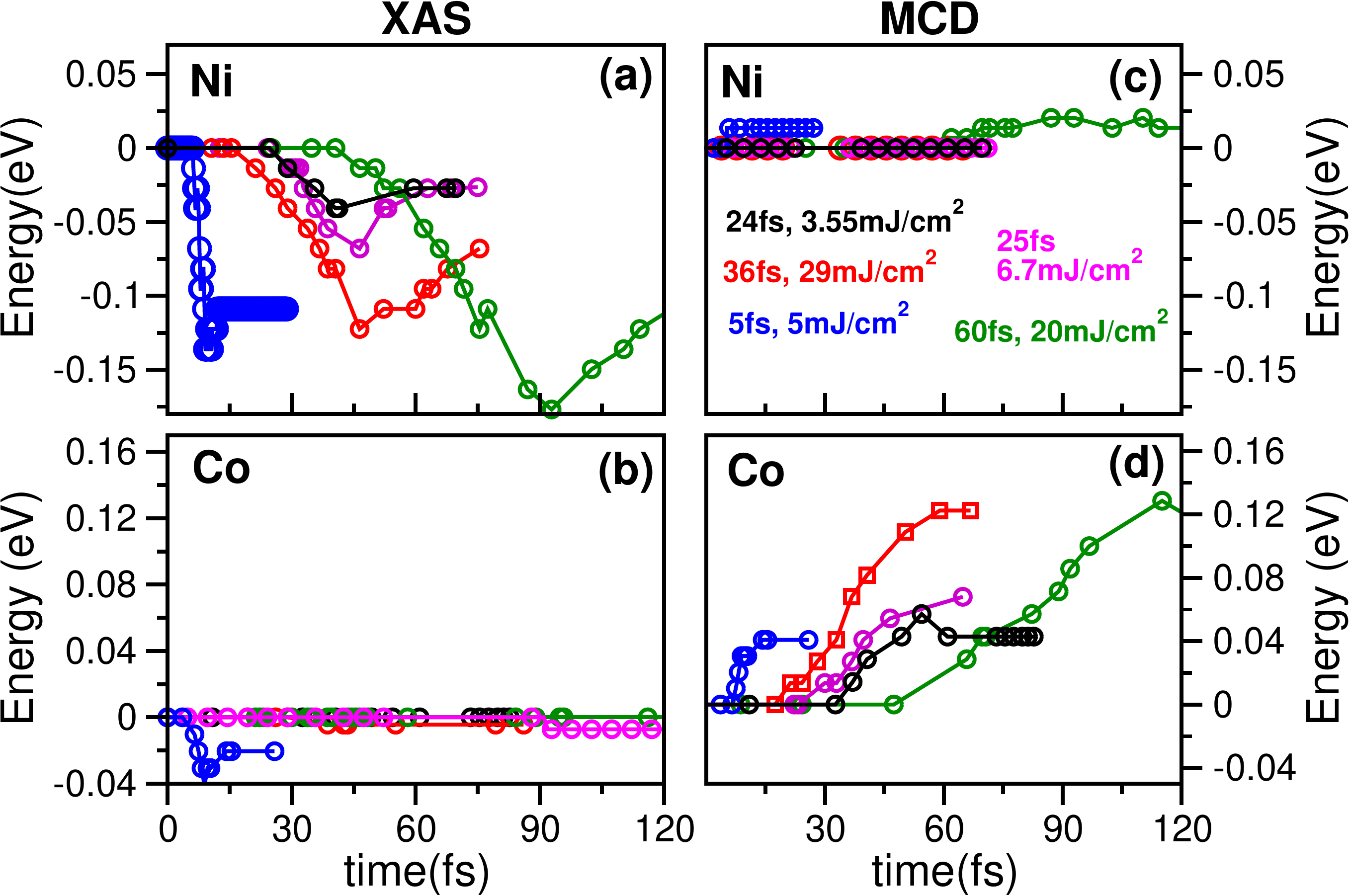}
\caption{Dynamics of L3 peak position upon laser pumping: XAS L3 peak for (a) Ni and (b) Co. MCD L3 peak for (c) Ni and (d) Co. The linearly polarized pump laser pulses (with polarization parallel to the spin-quantization axis) have a central frequency of 1.55~eV, with the duration and incident fluence as shown in the legend. The Ni XAS and Co MCD L3 peaks show a shift independent of the pulse parameters, however the amount of the shift depends strongly upon the pump pulse. In contrast, the Ni MCD and Co XAS peaks display almost no shift, a fact that holds for all pulse parameters.
}\label{fig:ps}
\end{figure}

\begin{figure}[t!]
\includegraphics[width=\columnwidth, clip]{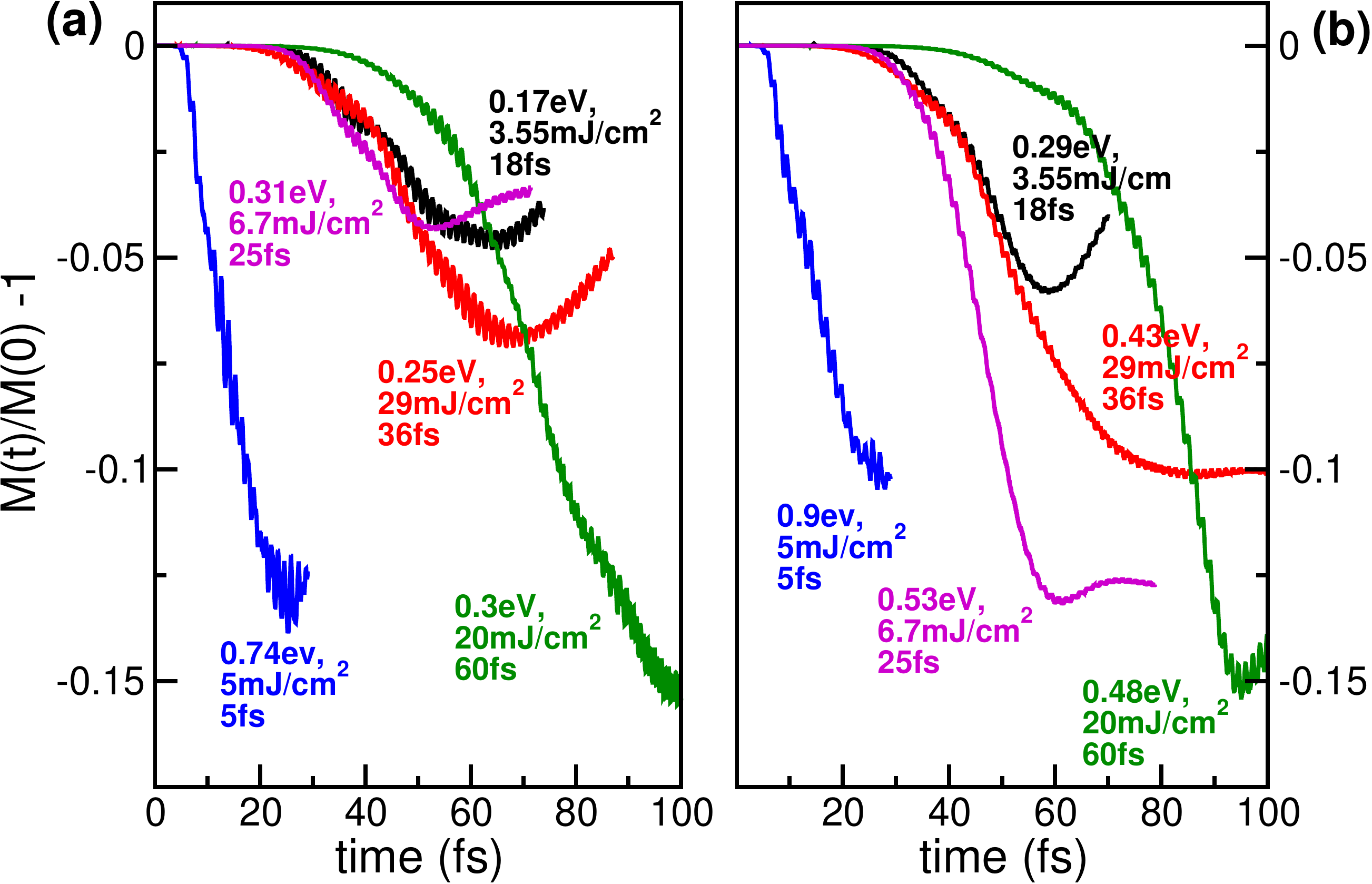}
\caption{Normalized change in moment as a function of time in laser pumped (a) Ni and (b) Co. The central frequency of all the pulses is 1.55~eV. The duration (in fs), the incident fluence (in mJ/cm$^2$), and the absorbed energy per unit-volume (in eV) are quoted in the legend. For all cases these pump pulses are linearly polarized with polarization parallel to the spin-quantization axis. As in experiments, we find that for a given absorbed energy, Ni demagnetizes more that Co.}\label{fig:pm}
\end{figure}

\emph{Dependence of the peak shifts on laser pulse parameters}: Thus far we have probed the shifting of spectral peaks in the XAS and MCD via a single set of laser pulse parameters, and we thus now examine the impact of the form of the laser pulse on the shifting of the XAS and MCD L edge peaks. In Fig.~\ref{fig:ps} is shown the XAS and MCD peak shift in Ni and Co, for a wide range of pulse parameters. As can be seen for Ni in each case the XAS L3 peak shifts downward in energy (see Fig. ~\ref{fig:ps}a), by up to 150~meV, while the MCD L3 peak (see Fig. \ref{fig:ps}c) shows almost no shift (maximum being 20~meV). In Co MCD L3 peak (see Fig. \ref{fig:ps}d) shifts always upwards in energy, by up to 120~meV, while the XAS L3 peak (see Fig. \ref{fig:ps}c) shows very small downward shift, with a maximum value of 40~meV. The amount of the shift is dependent upon the laser pulse parameters.
The magnitude of the XAS peak shift in Ni corresponds very closely to the size of the laser induced demagnetization, as may be seen by comparison of the peak shift (Fig.~\ref{fig:ps}a) and normalized moment loss (Fig.~\ref{fig:pm}a). In the case of the MCD shift in Co such a clear link is not seen, with the loss of moment quite different for laser pulses that lead to similar peak shift. This corresponds with the more complex underlying electronic behaviour: the dramatic skew about $E_F$ in moment loss is a result both of (i) optical (spin preserving) transitions from minority below to above $E_F$ as well as (ii) spin-orbit induced transitions from majority into minority below $E_F$. The skew of moment loss about $E_F$ will evidently depend on how these two processes compensate each other, which is expected to be rather sensitive to pulse parameters.

It is important to note that for a given absorbed fluence Ni demagnetises more than Co. However, we find that for a fixed pulse parameters Co absorbs more than Ni. This is due to non-linear effects that result in excitations far above the Fermi level, up to 6~eV above the Fermi level as shown in a recent experimental investigation\cite{Sidiropoulos2021}. Such states, however, have very short lifetime and thus quickly decay to lower energy states by scattering events, resulting in additional transitions to lower energy states. This decay of high energy states is not included in our theoretical approach. We can, however, determine the energy "locked in" to these high energy states\cite{Pellegrini22} and find it to be of the order of 80\% of the incident energy. Thus in order to reproduce the energy available for creating low energy excitations in experiment, theoretically we require a significantly higher absorbed fluence to compensate for the energy "locked into" these high energy states. In Co the empty $d$-band width is much larger than Ni and hence this effect is much more pronounced leading to the higher absorbed energy.

In this work, we have examined the origin of the shift downwards in energy of the XAS L3 edge peak in Ni, and the absence of a corresponding shift in the MCD L3 edge peak. Using first principles simulations of ultrafast demagnetisation we are able to reproduce this effect and, moreover, predict a strikingly different behaviour will be found in Co with for this material the MCD peak shifting significantly upwards in energy with only negligible change in the XAS peak position.

While in the past diverse explanations have been put forward for the peak shift in Ni, our work establishes a purely electronic origin for this effect. In Ni the nearly full $d$-band restricts the energies above $E_F$ into which charge can be excited, resulting in a charge excitation skewed to negative energies and a shift downwards of the XAS peak. In Co, in contrast, the less full $d$-band results in more evenly distributed excitation of charge from below to above $E_F$ suppressing the peak shift. Spin orbit induced transitions from majority to minority channels below $E_F$ result in a very small change in moment below $E_F$, with the demagnetisation driven by a gain in negative moment (i.e. minority charge) above $E_F$. In Ni this is restricted to be close to $E_F$, yielding no MCD peak shift, whereas in Co this extends to up to 2~eV above $E_F$ resulting in a significant shift of the MCD peak.

We close the paper by summarising our three key predictions: (i) The XAS L3 edge peak shift found in Ni will be much suppressed in Co, with instead a pronounced shift in the dichroic L3 edge signal, (ii) as these peak shifts are driven by laser induced changes in band occupation, they will occur for ultrafast pulses below the time scale of other excitations (e.g. magnons), and (iii) the magnitude of the shift will be dependent upon the laser pump pulse parameters with the XAS shift approximately proportional to the change in normalized moment, with a more complex dependence for the MCD peak shift in Co. Fully energy- and time- dependent experiments will be able to confirm the predictions made in the present work. 

\emph{Acknowledgements}: Sharma, JKD, CvkS and SE would like to thank the DFG for funding through project-ID 328545488 TRR227 (project A04 and A02). Shallcross would like to thank DFG for funding through SH498/4-1. The authors acknowledge the North-German Supercomputing Alliance (HLRN) for providing HPC resources that have contributed to the research results reported in this paper.

%

\end{document}